MACHINE LEARNING MODELING OF siRNA STRUCTURE-POTENCY
RELATIONSHIP WITH APPLICATIONS AGAINST SARS-CoV-2 SPIKE GENE

By

Damilola Oshunyinka

A Thesis Submitted to

The Faculty of the Graduate School at

North Carolina Central University

In Partial Fulfillment of the Requirements

For the Degree

Master of Science

Durham

2023

Approved by

_________________________

Committee Chair

_________________________

Committee Co-Chair

_________________________

Committee Co-Chair

ABSTRACT

DAMILOLA OSHUNYINKA., M.S. Machine Learning Modeling of siRNA Structure-Potency Relationship with Applications Against SARS-CoV-2 Spike Gene. (2023)
Advisor: Dr. Weifan Zheng

The pharmaceutical Research and development (R&D) process is lengthy and costly, taking nearly a decade to bring a new drug to the market. However, advancements in biotechnology, computational methods, and machine learning algorithms have the potential to revolutionize drug discovery, speeding up the process and improving patient outcomes. The COVID-19 pandemic has further accelerated and deepened the recognition of the potential of these techniques, especially in the areas of drug repurposing and efficacy predictions. Meanwhile, non-small molecule therapeutic modalities such as cell therapies, monoclonal antibodies, and RNA interference (RNAi) technology have gained importance due to their ability to target specific disease pathways and/or patient populations.

In the field of RNAi, many experiments have been carried out to design and select highly efficient siRNAs. However, the established patterns for efficient siRNAs are sometimes contradictory and unable to consistently determine the most potent siRNA molecules against a target mRNA. Thus, this paper focuses on developing machine learning models based on the cheminformatics representation of the nucleotide composition (i.e. AUTGC) of siRNA to predict their potency and aid the selection of the most efficient siRNAs for further development. The PLS (Partial Least Square) and SVR (Support Vector Regression) machine learning models built in this work outperformed previously published models. These models can help in predicting siRNA potency and aid in selecting the best siRNA molecules for experimental validation and further clinical development. The study has demonstrated the potential of AI/machine learning models to help expedite siRNA-based drug discovery including the discovery of potent siRNAs against SARS-CoV-2.

## DEDICATION

I hereby dedicate this project to Yahweh for the wisdom, intellect, and courage to successfully complete the project and write this paper. Also, I dedicate this work to my family (Daniel, Sis Mo, Anjola, Bukunmi, Aanuoluwakiisha Oshunyinka) and my friends (Glory, John, Abisola, Kenny, Srivalli) for their moral support throughout the course of the project.

ACKNOWLEDGMENT

This endeavor would not have been possible without the thorough supervision, guidance, and support of my academic supervisor, Dr. Weifan Zheng, and co-supervisor Dr. Xialang Dong. Many thanks to my committee members for their constructive feedback. Special thanks to the entire faculty of the Department of Pharmaceutical Sciences, Biomanufacturing Research Institute and Technology Enterprise.

# TABLE OF CONTENTS

LIST OF FIGURES

LIST OF TABLES

## ABBREVIATIONS

SVR – Support Vector Regression

PLS – Partial Least Square

siRNA – Small or short interfering RNA

ANN – Artificial Neural Network

SARS-CoV-2 - Severe Acute Respiratory Syndrome Coronavirus 2

BCUT - Burden, Connectivity, and Topology descriptors

ML – Machine Learning

RMSE – Root Mean Square Error

RISC - RNA-induced silencing complex

ACE2 – Angiotensin Converting Enzyme-2

$r$ – Pearson correlation coefficient

$r^2$ – Coefficient of determination

CHAPTER I

INTRODUCTION

Overview of Drug Discovery Process

It is well established that the traditional model of R&D is synonymous with a high-cost and time-consuming drug discovery and development process. Recent analysis shows drug discovery and development process takes over $1 billion from initial research to the finished and approved marketable product [1-3]. In parallel, the ratio of new drug applications in comparison to the billions spent in R&D is endemically in decline suggesting limitations in the traditional R&D model [4]. Efforts to curb R&D expenses and save time have included the adoption of computational approaches through virtual screens of potential hits, optimization of leads, and implementation of in-silico tools for drug repurposing and prediction of ADMET (absorption, distribution, metabolism, excretion, and toxicity) properties [3, 5, 6]. Beyond the application of computerized systems to biological data in molecular docking, and pharmacophore modeling, computational algorithms have covered all aspects of the drug discovery process [7]. Big Data generated from high throughput sequencing, high throughput screening, large databases like PubChem and ChEMBL[8] has encouraged the adoption of computational techniques including machine learning algorithms to make predictions, classify, and draw inferences from the complex and voluminous biological data [9-12].

<u>Novel Biotechnologies to Drug Discovery</u>

The incorporation of translational research as an approach to drug discovery has paved the way for the emergence of new technologies in disease treatment including CAR-T-Cell therapy, the use of monoclonal antibodies, siRNA, and other gene therapy applications [13]. Early on, drug discovery stemmed from the isolation of active compounds from medicinal plants or dyes such as morphine from opium poppies and quinine from cinchona bark, leading to the development of modern pharmacology [14]. In the early 20th century, synthetic small-molecule chemicals began to be used as drugs, including aspirin and antibiotics. Many of these drugs are still widely used today and include many classes of drugs, such as statins, antihypertensives, and antidepressants [15]. Advancements in biochemistry and molecular biology including the Human Genome Project coupled with the development of bioinformatics tools has contributed to elucidating the genetic basis of diseases, opening better avenues to discover drug targets, and leading to the development of biologics including monoclonal antibodies, growth factors, and hormones used in the treatment of cancer, autoimmune disorders, and diabetes. [16-18]. Further, technological advancements such as the three-dimensional characterization of biological macromolecules using high-field nuclear magnetic resonance spectrometers, PCR-based recombinant DNA technology, cryogenic crystal handling, and high-speed computing make huge contributions to the defining molecular targets. Corresponding to these technological advancements, protein drug - protein target interactions exist to provide adequate therapies where small-molecule drugs have failed [19]. In recent years, advances in gene editing technology, such as CRISPR/Cas9 and nucleic acid-based therapy such as antisense oligonucleotides and RNAi have allowed for precise modifications to the genome [20, 21].

<u>Machine Learning Technologies</u>

Machine learning which involves both supervised, semi-supervised and unsupervised learning methods can and has been employed in all aspects of drug discovery and development processes [7]. The machine learning method mostly relies on experimental data which serves as input training data for the machine learning algorithms to learn the pattern of relationships [22, 23]. The input training data in supervised learning approach contains dependent and independent variables. The independent variables are the features that defines the outcome also known as the dependent variables or labels [24]. Regression and classification models are examples of supervised learning methods. An unsupervised learning approach is used on dataset with no labels. Clustering and principal component analysis models are examples of the unsupervised learning approaches [24]. The semi-supervised machine learning approach employs a bit of both the unsupervised and supervised models. The features of a dataset generally determine the choice of machine learning models to deploy. Regressions models are typically applied to numerical values and classification models to categorical values [22]. Deep learning (DL) a subset of machine learning that uses neural networks (Figure 1) with many layers to imitate human thinking and learn the complex relationships in data [25] has been successfully used to predict protein structures, drug toxicity, drug-target interactions, design new chemical entities, and facilitate drug repurposing by the therapeutic classification of drugs based on transcriptomic profiles [10, 26-28]. Unsupervised machine learning algorithms has also been deployed to predict drug-target relationships based on pharmacophore descriptors [29]. Antibacterial resistant microbes prompt the use of predictive artificial intelligence algorithms to discover new antibiotic agents where screening of chemical compound libraries couldn't yield results. The application of neural network models to predict antimicrobial candidates yielded the discovery of eight new antibiotics structurally different from existing antibiotics [30]. Machine learning has also been employed on data generated from genomic assays to predict splice sites, transcriptional promoters, enhancers and annotate genes [31-33]. A study done by Yan and colleagues reports the use of deep learning to predict differentiation potentials of mesenchymal stem cells via single cell RNA sequencing data to promote precision in cell therapy and regenerative medicine [34].

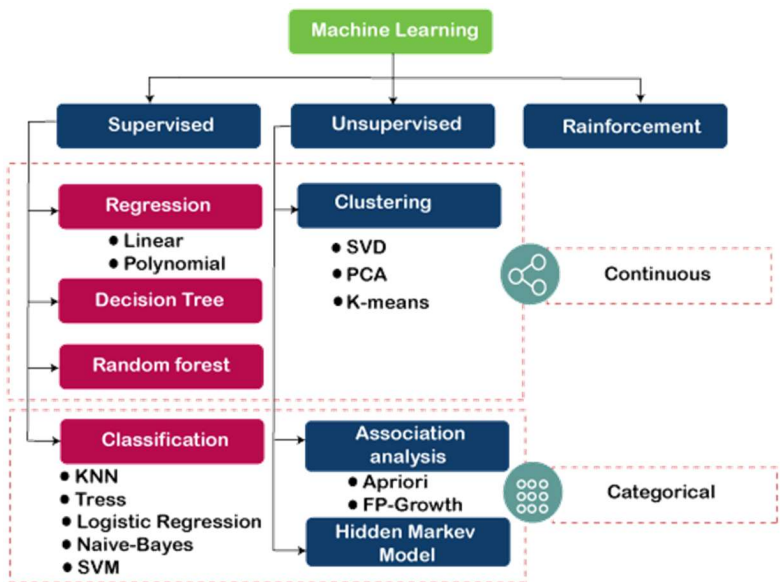

**Figure 1**. Chart of Machine learning algorithms. Source: JavaTpoint

COVID-19 Therapeutics

The severe acute respiratory syndrome coronavirus 2 (SARS-CoV-2) is a virus that causes COVID-19 (coronavirus disease 2019), a respiratory illness that first emerged in Wuhan, China in December 2019 and has since spread globally, resulting in a pandemic [35]. SARS-CoV-2 belongs to the family of coronaviruses, which also includes other viruses that cause respiratory illnesses such as SARS (severe acute respiratory syndrome) and MERS (Middle East respiratory syndrome). SARS-CoV-2 is a positive-sense single-stranded RNA virus that enters human cells through the ACE2 receptor and replicates in the respiratory tract, leading to symptoms such as fever, cough, and difficulty breathing [36]. The urgency to discover effective treatment for COVID-19 led to the emergency use authorization of several therapeutics. Antiviral drugs such as Remdesivir and Favipiravir inhibit viral replication by targeting the RNA-dependent RNA polymerase of SARS-CoV-2 [37, 38]. There are also combinations of monoclonal antibodies including Casirivimab/imdevimab and Bamlanivimab/etesevimab that target the spike protein of SARS-CoV-2 to prevent it from entering human cells.

During the heat of the SARS-CoV-2 pandemic, Ribonucleic Acid interference (RNAi), a gene-silencing technology against RNA viruses was actively pursued as a treatment option against COVID-19. The S gene sequence of SARS-CoV-2 codes for the spike protein that binds the virus to angiotensin-converting enzyme2 (ACE2) as a receptor, thus facilitating entrance into human cells [39]. Silencing the SARS-CoV-2 S gene using RNAi mechanism prevents the translation of spike glycoprotein used by the virus in cellular invasion [40]. Chen and colleagues performed one of the earlier studies adopting RNAi as a treatment option for COVID-19 [40]. They proposed a Simple Multiple Rules Intelligent method (SMRI) which incorporates machine learning and statistical methods built on the thermodynamic BPTC (Base Preference and Thermodynamic Characteristic) parameters and secondary features of the small interfering RNAs (siRNAs) as the basis for the high silencing efficiency in siRNAs against SARS-CoV-2's Spike gene. To filter out ineffective or non-synthesizable siRNAs from the predictions, the SMRI method incorporated a siRNA Extended Rules (SER) index which combines the summary of rules defining highly efficient siRNAs previously proposed by Reynolds and colleagues [41], Amarzguioui and Prydz [42], Hsieh and colleagues [43]. The resulting predicted silencing efficiency matched those designed by statistical analysis of empirical rules.

RNAi Therapeutics

Since the discovery of RNAi in 1998 by Andrew Fire and Craig Mello [44], the application of the underlying principles of RNAi has provided the basis for gene function studies, disease pathway analysis, and therapeutics for gene defects, cancers, and autoimmune disorders via sequence-specific gene silencing technology [45-47]. Fire and Mello were awarded a Nobel Prize in Physiology and Medicine in 2006 for their discovery. They observed that small RNA molecules produced from double-stranded RNAs (dsRNA) induced potent and specific gene-silencing in nematode Caenorhabditis elegans. These small RNA molecules were later regarded as small or short interfering RNAs (siRNAs). Like other antisense strategies including antisense oligonucleotides and ribozymes, RNAi seeks to silence genes by degrading target mRNA but by using small interfering RNAs (siRNAs), microRNAs (miRNAs), and short hairpin RNAs (shRNAs) [47]. However, RNAi is more potent and efficient than other antisense strategies [48].

Mechanism of siRNA Gene-silencing

Short-interfering RNAs (siRNAs) are a type of RNA enzymatically cleaved from double-stranded RNAs (dsRNAs) and are responsible for the gene-silencing observed in RNA interference (RNAi) [49-51]. The mechanism of gene-silencing existed prior in plants as a process of co-suppression and in fungi as a process of quelling [52]. In co-suppression, the introduction of a transgenic copy of an endogenous gene causes the silencing of both genes [53, 54] whereas in fungi, quelling is usually triggered by the presence of transposons or other repetitive DNA sequences within the genome [55-57]. Both naturally occurring processes and the RNAi pathway found in several eukaryotic organisms represent genome defense mechanisms against transposons and viruses as well as regulate cellular, physiological processes [55, 56].

The endogenous RNAi pathway which occurs predominantly in the cytoplasm is triggered by dsRNA which can be exogenously introduced as bimolecular duplexes or short hairpins (shRNAs) via plasmids or viral vectors (Figure 2) [58-60]. Exogenously introduced dsRNAs are cleaved into siRNAs by a Dicer, a member of the RNase III family of nucleases [49]. These siRNAs, 21 - 23 nucleotide (nt) base-pairs in length with overhangs on 3′ ends, are conjugated to the RNA-induced silencing complex (RISC). The RISC guides single strands of siRNA based on complementarity to target mRNA for cleavage [50, 61]. The mRNA is cleaved at 21-23nt intervals and broken down by cellular exonucleases [40, 62].

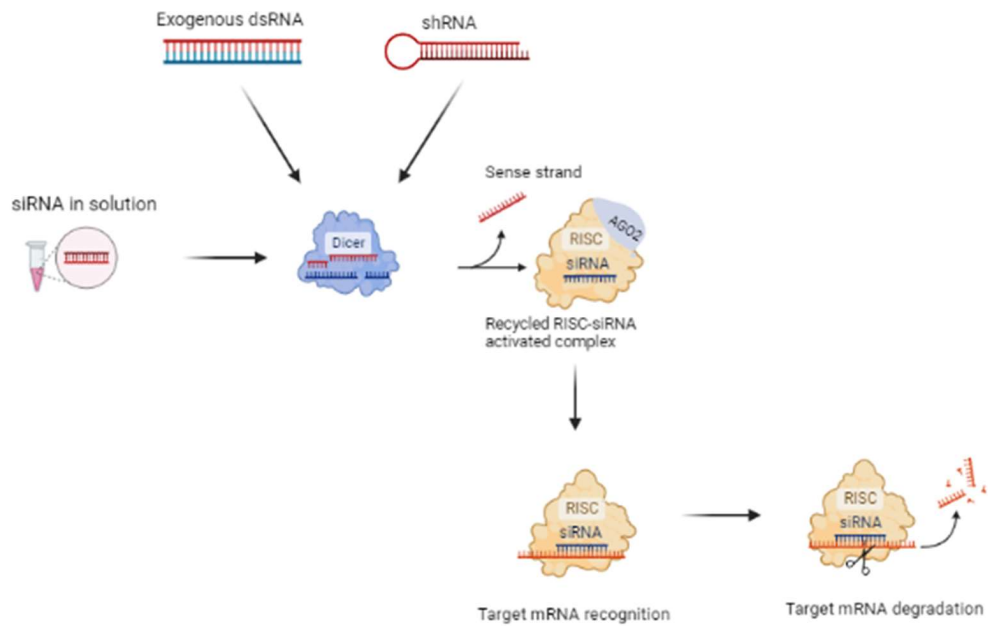

**Figure 2**: Schematic diagram of siRNA-mRNA degradation. Exogenous dsRNA and shRNA or synthetic siRNA introduced into the cell cytoplasm is cleaved by Dicer into fragments. The antisense strand conjugated with the RISC recognizes target mRNA and leads to target cleavage. The sense strand is further cleaved by exonucleases. Created using BioRender.

DsRNA also forms endogenous dsRNA hairpins when adjacent complementary segments in endogenous RNA transcripts fold back on themselves. The dsRNA hairpin is further cleaved by Dicer to siRNAs or microRNAs (miRNAs) [63]. Bartel, in his review of the biogenesis of miRNAs [59] also describes the role of the endonuclease, Dicer, in the maturation of miRNA another subtype of dsRNAs employed in the regulation of transcriptional gene expression. After the transcription of primary miRNA (pri-miRNA) from DNA in the cell nucleus by the enzyme RNA polymerase II, a complex of proteins which includes the enzyme Drosha processes the pri-miRNA to form a shorter hairpin-shaped precursor miRNA (pre-miRNA). The pre-miRNA gets transported out of the nucleus and into the cytoplasm, where it is further processed by Dicer, along with other proteins, to form a mature miRNA as shown in Figure 3. siRNAs are also introduced directly with high potency gene-silencing effect but their transient effect means repeated administration unlike the administration of a single viral vector of dsRNA or shRNA [47].

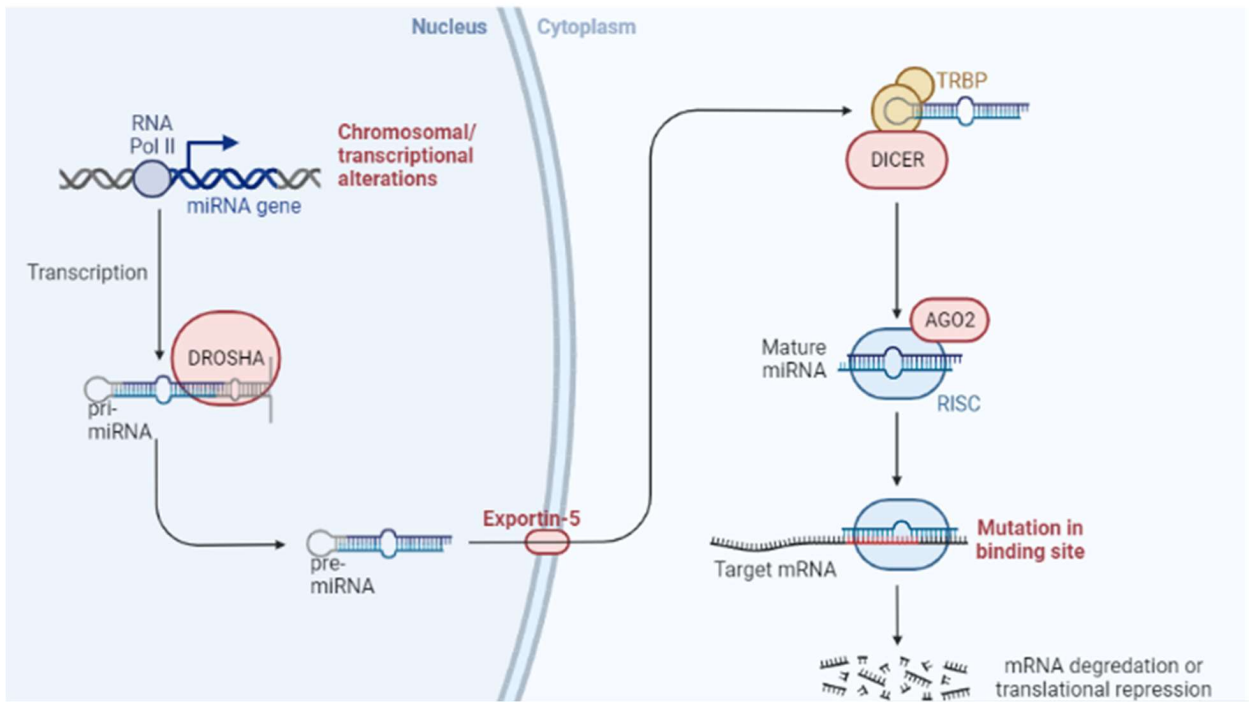

**Figure 3**: Schematic diagram of miRNA-mRNA target degradation. Created with BioRender.

Both siRNA and miRNA fragments of dsRNAs assemble into ribonucleoprotein effector complexes which contain the highly conserved Argonaute proteins (AGO2) that cleaves one (passenger or sense) strand of the siRNA allowing the incorporation of the other (guide or antisense) strand into the RISC [61, 64, 65]. The choice of which strand becomes the passenger or guide strand is a result of thermodynamic instability found at the 5′-anti-sense (AS) terminal regions of functional siRNAs [61, 66]. The common method to experimentally determine gene silencing efficiency of siRNAs is using reporter assays where target mRNA levels are quantified after transfecting cell lines of interest with siRNA [66-68].

<u>Current and past therapeutic application of siRNAs</u>

The high specificity found in siRNA-induced gene silencing where mRNA is degraded based on siRNA nucleotide base complementarity provides new therapeutic methods beyond the standard use of small molecules, especially for targets that are not amenable to traditional therapeutic approaches [45, 46, 69]. One challenge to the development of therapeutic siRNAs is systemic delivery. Naked, unmodified siRNAs are known to be unstable, have low plasma half-life, lack drug-like properties, and are unable to cross systemic cell membranes due to negative charges [46, 70]. To overcome these challenges, several delivery vehicles, chemical modification, and conjugation methods have been explored to confer drug-like properties on siRNAs and allow efficient delivery of siRNA to systemic targets including conjugation with lipids, modification with 2′-O-methyl or 2′-fluoro modifications, GalNAc (N-acetyl galactosamine)–siRNA conjugates, nanoparticles, and liposomes. At present, there are more than 7 siRNA therapeutics in phase III clinical trials with several more in other clinical trial phases targeted against hemophilia, cancers, metabolic, ocular, and infectious diseases [71-73]. ONPATTRO® (patisiran) and GIVLAARI™ (givosiran) applied to the treatment of rare genetic disorders are examples of FDA-approved siRNA therapeutics [71, 74]. Experimental results also show that siRNAs can reduce the viral genome replication and expression of highly conserved regions of SARS-CoV-2 [68, 75-77]. Although, clinical trials studying the efficacy of siRNA- based therapeutics in SARS-CoV-2 are yet to commence [78].

Other challenges associated with the development of siRNAs as a therapeutic are "off-target" effects where siRNA regulates the expression of unintended mRNAs [79, 80]. The unintentional uptake of the passenger (sense) strand into the RISC instead of the guide (antisense) strand is another cause of off-targeting effects [79]. To prevent off-targeting, most experimental design for effective siRNAs is coupled to a BLAST (Basic Local Alignment Search Tool) analysis and Smith-Waterman search for homogenous sequence analysis with human genome to screen off siRNAs with perfect or almost perfect complementarity with unintended targets [40, 78, 81, 82]. siRNA sequences with at least three mismatches are retained and regarded as safe for human use [40].

## History of siRNA Design (Rule-based and computational)

The high specificity found in siRNA-induced gene silencing means observation of thorough details in the design of siRNA to avoid 'off-targeting' effects or the design of inefficient siRNAs. A single base difference between the antisense strand of an siRNA and its target mRNA substantially reduces the gene-silencing efficiency of the siRNA [40]. Another studies demonstrates that as little as 11 matches between siRNA and unintended targets may cause off-target silencing [83].

Since the discovery of the technology of RNAi via siRNAs, the design of effective siRNAs has progressed from the use of rational rules to computational algorithms, in a bid to improve the selection of potent and highly efficacious siRNA [80]. In a result comparing the thermodynamic properties of functional and non-functional siRNAs,

Khvorova and colleagues described that at the 5′-anti-sense (AS) terminal regions, functional siRNAs showed more thermodynamic instability than functional siRNAs. As demonstrated in their experiment, strands with low internal stability at the 5′ end are retained within the RISC for mRNA cleavage [66]. From the systematic analysis of 180 siRNAs, Reynolds and colleagues determined certain characteristic features of functional siRNAs [41]. Results of their analysis showed that functional siRNAs were thermodynamically unstable at the 3′-terminus of the sense strand, had lower presentations of G-C base pairs, and specific nucleotides motifs are preferred in specified positions [41]. The presence of motifs such as a stronger G-C base pairing at position 1, a weaker A-U base pairing at position 19 are characteristic of functional siRNAs. The application of these features into an algorithm improved the selection of potent siRNAs, though false negatives were also identified. In another publication, Elbashir and colleagues in describing the selection of potentially effective siRNAs demonstrated that 5′ or 3′ untranslated regions and regions close to the start codons of mRNA should be avoided when designing siRNAs. Specific sequences in the mRNA are considered first for potential siRNA duplex selection before other regions in the mRNA. There is also a preference of uridine residues as the nucleotide overhangs in the siRNA duplex (Elbashir et al., 2002). Effective siRNAs reportedly have more A or U nucleotides at the 5′ end of the antisense (guide) strand as opposed to the G or C nucleotides found at the 5′ end on the antisense strand in ineffective siRNAs. In contrast, the 5′ terminal of sense (passenger) strand in highly effective siRNAs have more G or C nucleotides, while A or U nucleotides are presented more at this strand position ineffective siRNAs [67]. Highly effective siRNAs also boasted of A or U nucleotides in four out of seven nucleotides present at the 5′ terminal of the antisense strand. siRNAs rich in G or C nucleotides in this strand terminal are ineffective while siRNAs with mixture of this nucleotides have moderate silencing efficiency [42, 67]. The strategic position of these nucleotides is necessary for siRNA conjugation with the RISC and further supports hydrogen bonding with target mRNA [67]. Other studies emphasize the secondary structure of target mRNAs as a more important criteria for designing siRNA since they are essential for target recognition by siRNAs [84, 85].

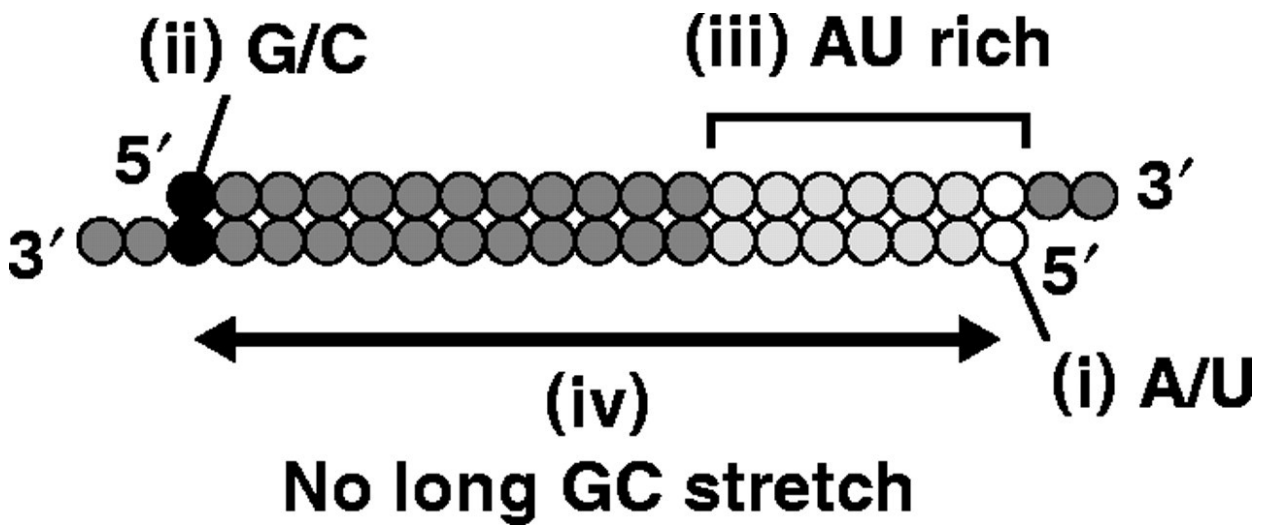

**Figure 4**: Common features of highly effective siRNAs. Source: Yuki Naito and colleagues [86].

The design of siRNAs based on each rule produced inefficient results, hence, some of these rules are incorporated into algorithms and have proven effective in the design of siRNAs against new targets. Each rule is assigned a score and siRNAs are scored based on these assigned scores, siRNAs passing a threshold score are preferentially selected as effective [42, 82]. The rules can be differentiated into specifics mandatory for functional siRNAs while others are basis to improve siRNA efficiency[82]. Some siRNA design algorithms are built on contradictory rules, each rule with predefined parameters that provides users inflexibility with choice of rules as needed for their experiments [82]. While these rules help distinguish potentially effective from ineffective siRNAs, there are inconsistencies, report of false negatives and no report of siRNA efficacy data to determine the efficiency of the rules [87]. Importantly, the rule-based selection of functional siRNAs is not sufficient to account for the properties of an effective siRNA and inform the design of the most potent and efficacious siRNA against a single mRNA target [88]. Beyond the classic rules of siRNA design (Figure 4), advanced statistical and machine learning methods including supervised learning support vector machine (SVM), artificial neural networks are so employed to design siRNAs [88]. Peek Andrew in a study applying SVM for siRNA predictions mapped features of corresponding siRNAs such as thermodynamics of stability energies, positional motifs, guide strand, and target mRNA strand secondary structures, into vectors. Each feature was tested for correlation to siRNA activity using Correlation based Feature Selection (CSF) method [89] to filter a subset of features relevant to siRNA activity for SVM modeling [88]. Selecting only subset of features improved the performance of the SVM model. Guangtao Ge and colleagues trained a Bayesian neural network model on 180 siRNAs with experimental determined potencies [90]. Their model was built on RNA structure thermodynamic parameters. Huesken and colleagues [91] trained artificial neural networks (ANN) with siRNA sequences and their potencies experimentally determined from reporter assays as inputs. In contrast to previous approaches that trained machine learning algorithms on thermodynamics, strand structures, free energies, Zheng and Dong [92] applied the more statistical Partial Least Squares (PLS) method to build models that can predict siRNA potencies based on 12 cheminformatics (BCUT) descriptors including charge distribution, hydrophobic and polarization properties of individual nucleotide components of the siRNAs. The dynamics embedded in the cheminformatics descriptions of siRNA nucleotide compositions is expected to play a role in gene-silencing efficiency of the siRNA molecule. Results from this experiment was comparable to Huesken and colleagues, considering both machine learning models were built on the same siRNA input data but with different features [91, 92].

Specific Aims of this Study

Building up Zheng and Dong's experiment as described above, this present work aims to:

- Demonstrate the method of computationally predicting siRNA potencies based on 206 molecular descriptions of each nucleotide in the siRNA.
- Apply PLS, SVM machine learning algorithms to create models from learning the pattern of relationship between molecular descriptor features of 2,431 siRNAs, and their predetermined experimental potencies.
- Validate predictive models by application to predict the potencies of all potential siRNAs against SARS-CoV-2 spike gene based on their molecular description features. The validation of our model will provide a new alternative approach to designing and selecting highly potent siRNAs. Although computationally determined potent siRNAs would be experimentally validated, the experimental screening is streamlined to accommodate siRNAs with higher potencies to save time and expense.

CHAPTER II

MATERIALS AND METHODS

The workflow (Figure 5) describes the method used to run the computational analysis.

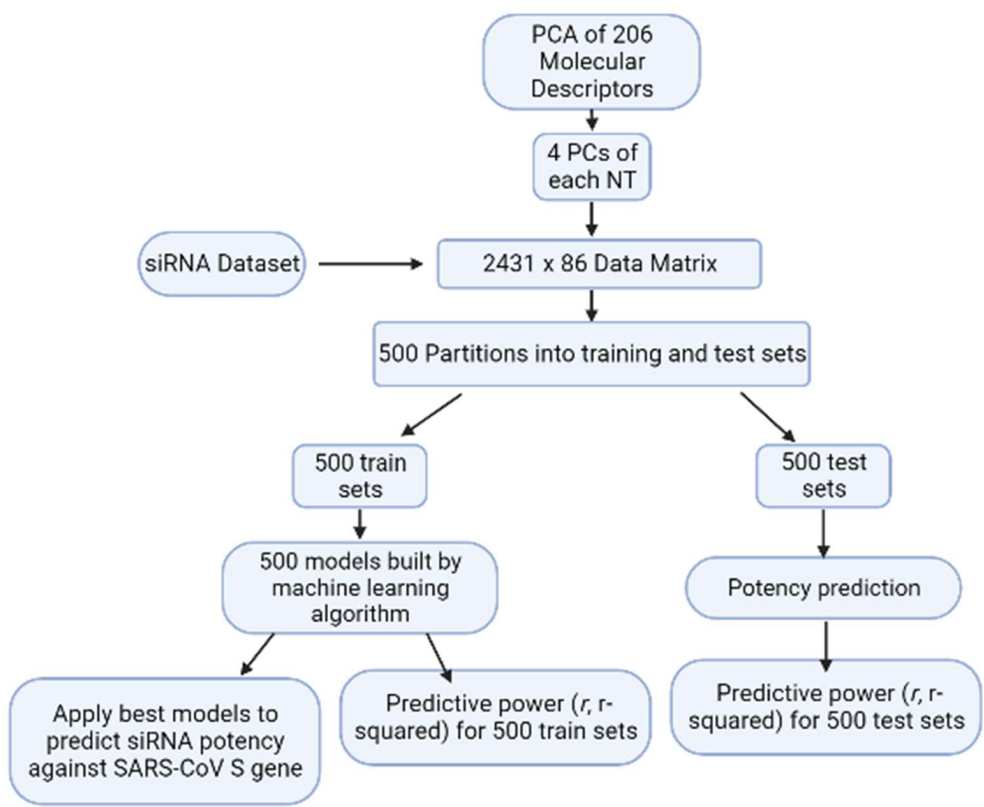

**Figure 5**: Workflow for computational modeling and application against SARS-CoV-2 gene. Created with BioRender.

The Huesken Dataset

2,431 siRNA antisense sequences were acquired from a genome-wide siRNA library designed against different human and rodent genes by Huesken and colleagues [91]. The Huesken dataset is the largest currently known public comprehensive library of siRNAs whose potencies were measured in a single experiment and is often used to develop and validate models, tools for siRNA design. Huesken and colleagues experimentally created and validated siRNA against 34 different mRNAs and their respective potencies. Luciferase reporter assays were used in the experimental analysis of siRNA gene-silencing efficiency where fusion of siRNAs with fluorescent protein reporter gene plasmid inserts bearing target cDNA reduces the reporter gene expression levels. The inhibition of luciferase activity was then measured, and the inhibitory activity of each siRNA was quantified. The resulting potency profile was validated with good Pearson correlation coefficient ($r \in$ [0.83, 0.89 0.66]) by comparison with the potency profile obtained from siRNAs targeting endogenous mRNA expression of human genes. The downregulation of mRNA expression was quantified using quantitative reverse transcriptase and western blot analysis. This validated the use of reporter assays to curate the vast siRNA library that offers comparison of siRNA potencies within a single assay.

Molecular Descriptors

The molecular operating environment (MOE) software developed by the Chemical Computing Group (Montreal, CA, USA) is a computer-aided molecular design platform predominantly used in drug discovery. Molecular descriptors are numerical vectors generated by algorithms based on structural, physiochemical, and geometrical properties of compounds. Line notations such as SMILES [93] and INCHI [94] are the modes by which cheminformatics libraries generate these molecular descriptors. Embedded in the MOE software is a package that calculates the molecular descriptors of compounds based on the SMILES structures (Table 1). The basis for the characterization of biological molecules using quantitative descriptors (Figure 6) have previously been described [95]. Theoretically, descriptors incorporate physical and chemical properties such as lipophilic, steric, electronic, hydrogen bonding, and electrostatic interactions that account for the biological activity observed. The MOE software was used to calculate the molecular descriptors of each nucleotide (Table 2). Each nucleotide (A, T, G, C, U) had 206 molecular descriptors which were summarized into 4 principal components. Each siRNA sequence with 21nt had 84 principal components (21nt x 4 principal component of each nucleotide molecular descriptors) which served as the features of the siRNA sequence.

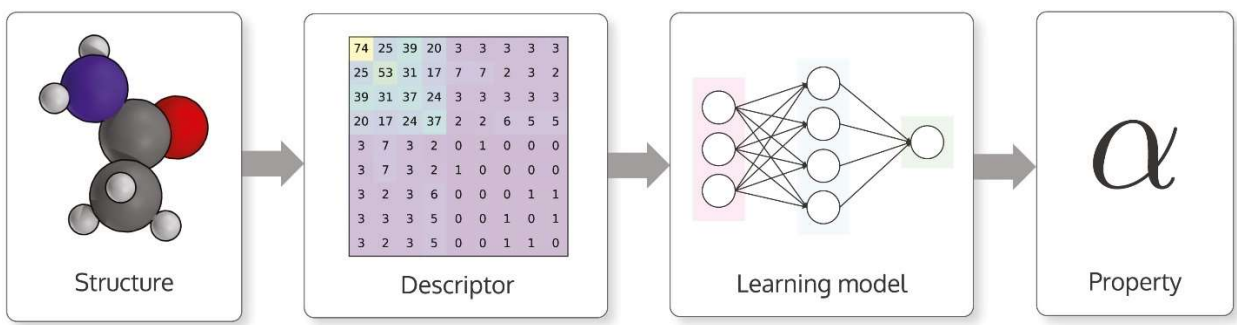

**Figure 6**. Molecular descriptor generation. Describes the workflow for generating descriptor properties from atomic structures of compounds. Structures are projected into vectors which are trained as inputs in a machine learning model to generate a molecular property as outputs. Source: DScribe: Library of descriptors for machine learning in materials science.

**Table 1**. Nucleotide-SMILES structures. SMILES structures are used to generate molecular descriptor properties on the MOE Software. Nucleotide structure and SMILES curated from <u>PubChem.</u>

| Nucleotide | Structure | SMILES | Descriptor Properties $(X_1, X_2, X\ldots, X_{206})$ |
|---|---|---|---|
| **A** | 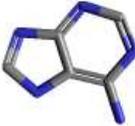 | C1=NC2=NC=NC(=C2N1)N | See Table 2. |
| **U** | 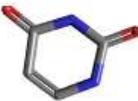 | C1=CNC(=O)NC1=O | |
| **G** | 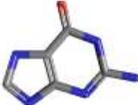 | C1=NC2=C(N1)C(=O)NC(=N2)N | |
| **T** | 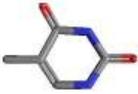 | CC1=CNC(=O)NC1=O | |
| **C** | 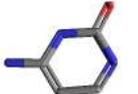 | C1=C(NC(=O)N=C1)N | |

Machine Learning Data Matrix

The data matrix is a fundamental concept in machine learning that represents the input data for a model. It is a two-dimensional array where each row corresponds to a sample, and each column corresponds to a feature or descriptor that characterizes the sample. The descriptors or attributes are the individual characteristics or properties that are used to describe each sample in the data matrix which can be numerical or categorical, continuous, or discrete depending on the data. The descriptor features used in this computational experiment to characterize the siRNA samples were obtained from the MOE software (Table 1 and Table 2). The 2,431 siRNAs, each with 84 molecular descriptors and their respective potencies made up a data matrix with 2,431 rows and 85 columns excluding the header row. 20% of this matrix was used as the test dataset while 80% was used as the training dataset.

Machine learning algorithms use the data matrix and descriptors to learn a function that maps inputs to outputs. These algorithms which can be supervised, unsupervised, or semi-supervised, are used for classification, regression, clustering analysis of the data matrix. There are several machine learning methods that map inputs to outputs including linear regression, logistic regression, support vector machines, decision trees, random forests, and neural networks. For this experiment, Support Vector Regression and the Partial Least Squares Regression analysis were machine learning methods applied to the data matrix.

Amongst other software tools employed to build Machine learning models, Python libraries which are software packages provide tools and functions to implement machine learning methods efficiently. Some of the most popular libraries for machine learning in Python are Scikit-learn, TensorFlow, Keras, PyTorch, and Pandas.

Predictive Modeling of siRNA Potency

The workflow for building predictive models using molecular descriptors on the Huesken dataset is shown in Figure 5. The steps involved includes:

(1) generating vector principal components from the principal component analysis of 206 molecular descriptors of each nucleotide that's composed in a siRNA using the MOE software. (2) using the Scikit-learn package [96] in python programming software to build up the matrix of the Huesken dataset which originally composed of 2431 rows and 3 columns (siRNA ID, siRNA sequence, siRNA potency) into a numerical representation of the siRNA sequence (2431 x 87 data matrix) by incorporating the 4 principal components of each nucleotide in the 21-nucleotide siRNA sequence (Figure 6);

(3) splitting the data matrix into 500 different random combinations of training and test sets.

(4) building predictive models from the training sets using SVR, PLS using the Scikit-learn package [96, 97] in python programming software.

(5) applying built models on test sets and calculating the predictive power (Pearson $r$, Root Mean Square Error (RMSE) values) for each model.

Application of the Machine Learning Models to SARS-CoV-2 Spike mRNA.

(6) generating siRNAs against 3,822-nucleotide long SARS-Cov-2 Spike gene.

(7) applying the best SVR, PLS and ANN models to predict the potency for siRNAs against SARS-Cov-2 Spike gene.

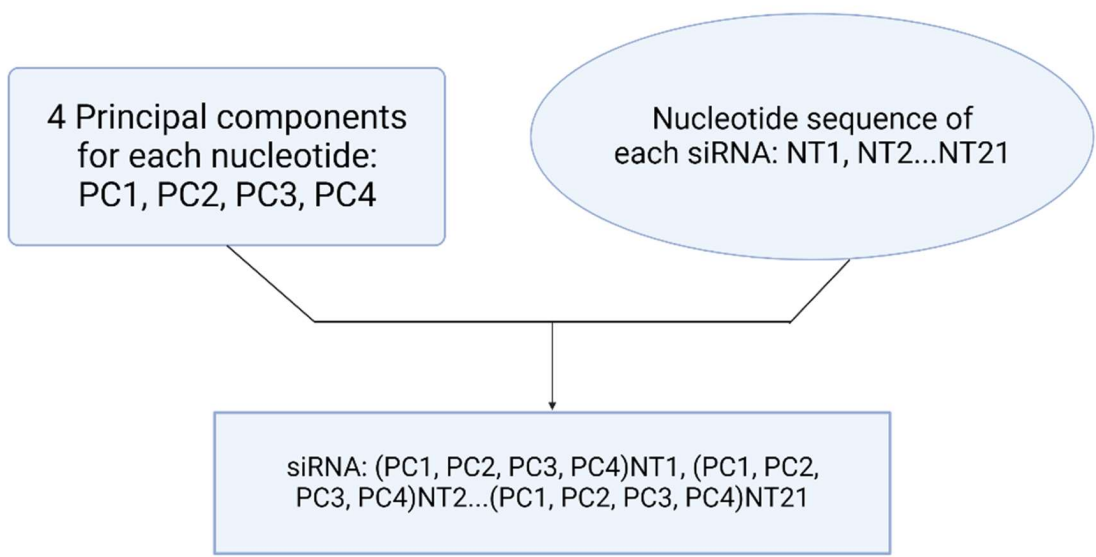

**Figure 7.** Data Matrix expansion. A Schematic illustration of building features of each siRNA by the principal components from the PCA analysis of 206 molecular descriptors of each nucleotide in the siRNA.

**Table 2.** MOE descriptors and definitions

| Descriptor | Description |
| --- | --- |
| VDistEq | If $m$ is the sum of the distance matrix entries then VdistEq is defined to be the sum of $\log_2 m - p_i \log_2 p_i / m$ where $p_i$ is the number of distance matrix entries equal to $i$. |
| VDistMa | If $m$ is the sum of the distance matrix entries then VDistMa is defined to be the sum of $\log_2 m - D_{ij} \log_2 D_{ij} / m$ over all $i$ and $j$. |
| b_1rotR | Fraction of rotatable single bonds: b_1rotN divided by b_heavy. |
| Weight | Molecular weight (including implicit hydrogens) with atomic weights taken from [CRC 1994]. |
| chi0 | Atomic connectivity index (order 0) from [Hall 1991] and [Hall 1977]. This is calculated as the sum of $1/\mathrm{sqrt}(d_i)$ over all heavy atoms $i$ with $d_i > 0$. |
| chi1 | Atomic connectivity index (order 1) from [Hall 1991] and [Hall 1977]. This is calculated as the sum of $1/\mathrm{sqrt}(d_i d_j)$ over all bonds between heavy atoms $i$ and $j$ where $i < j$. |
| VAdjEq | Vertex adjacency information (equality): $-(1-f)\log_2(1-f) - f \log_2 f$ where $f = (n^2 - m) / n^2$, $n$ is the number of heavy atoms and $m$ is the number of heavy-heavy bonds. If $f$ is not in the open interval (0,1), then 0 is returned. |
| VAdjMa | Vertex adjacency information (magnitude): $1 + \log_2 m$ where $m$ is the number of heavy-heavy bonds. If $m$ is zero, then zero is returned. |
| balabanJ | Balaban's connectivity topological index [Balaban 1982]. |
| PEOE_PC+ | Total positive partial charge: the sum of the positive qi. Q_PC+ is identical to PC+ which has been retained for compatibility. |
| PEOE_PC- | Total negative partial charge: the sum of the negative $q_i$. Q_PC- is identical to PC- which has been retained for compatibility. |
| PEOE_RPC+ | Relative positive partial charge: the largest positive $q_i$ divided by the sum of the positive $q_i$. Q_RPC+ is identical to RPC+ which has been retained for compatibility. |
| PEOE_RPC- | Relative negative partial charge: the smallest negative $q_i$ divided by the sum of the negative $q_i$. Q_RPC- is identical to RPC- which has been retained for compatibility. |
| mr | Molecular refractivity (including implicit hydrogens). This property is calculated from an 11 descriptor linear model [MREF 1998] with $r^2 = 0.997$, RMSE = 0.168 on 1,947 small molecules. |
| vsa_acc | Approximation to the sum of VDW surface areas of pure hydrogen bond acceptors (not counting acidic atoms and atoms that are both hydrogen bond donors and acceptors such as -OH). |
| vsa_don | Approximation to the sum of VDW surface areas of pure hydrogen bond donors (not counting basic atoms and atoms that are both hydrogen bond donors and acceptors such as -OH). |
| vsa_hyd | Approximation to the sum of VDW surface areas of hydrophobic atoms. |

| TPSA | Polar surface area calculated using group contributions to approximate the polar surface area from connection table information only. The parameterization is that of Ertl *et al.* [Ertl 2000]. |
|---|---|
| density | Molecular mass density: Weight divided by vdw_vol. |
| vdw_area | Area of van der Waals surface calculated using a connection table approximation. |
| vdw_vol | van der Waals volume calculated using a connection table approximation. |
| logP(o/w) | Log of the octanol/water partition coefficient (including implicit hydrogens). This property is calculated from a linear atom type model [LOGP 1998] with $r^2 = 0.931$, RMSE=0.393 on 1,827 molecules. |
| diameter | Largest value in the distance matrix [Petitjean 1992]. |
| radius | If $r_i$ is the largest matrix entry in row $i$ of the distance matrix $D$, then the radius is defined as the smallest of the $r_i$ [Petitjean 1992]. |
| weinerPath | Wiener path number: half the sum of all the distance matrix entries as defined in [Balaban 1979] and [Wiener 1947]. |
| weinerPol | Wiener polarity number: half the sum of all the distance matrix entries with a value of 3 as defined in [Balaban 1979]. |
| a_aro | Number of aromatic atoms. |
| b_1rotN | Number of rotatable single bonds. Conjugated single bonds are not included (e.g., ester and peptide bonds). |
| b_ar | Number of aromatic bonds. |
| b_double | Number of double bonds. Aromatic bonds are not considered to be double bonds. |
| rings | The number of rings. |
| zagreb | Zagreb index: the sum of $d_i^2$ over all heavy atoms $i$. |
| b_double/b_count | Number of double bonds. / Number of bonds (including implicit hydrogens). This is calculated as the sum of $(d_i/2 + h_i)$ over all non-trivial atoms i. |
| b_ar/b_count | Number of aromatic bonds / Number of bonds |
| b_single/b_count | Number of single bonds / Number of bonds |
| a_aro/a_count | Number of aromatic atoms / Number of atoms |
| a_don/a_count | Number of hydrogen bond donor atoms / Number of atoms |
| a_acc/a_count | Number of hydrogen bond acceptor atoms (not counting acidic atoms but counting atoms that are both hydrogen bond donors and acceptors such as -OH). / Number of atoms |
| a_hyd / a_count | Number of hydrophobic atoms. / Number of atoms |
| nX | Number of halogen atoms |
| nX/a_count | Number of halogen atoms / Number of atoms |
| Kier1 | First kappa shape index: $(n\text{-}1)^2 / m^2$ [Hall 1991]. |
| Kier2 | Second kappa shape index: $(n\text{-}1)^2 / m^2$ [Hall 1991]. |
| Kier3 | Third kappa shape index: $(n\text{-}1)\,(n\text{-}3)^2 / p_3^2$ for odd $n$, and $(n\text{-}3)\,(n\text{-}2)^2 / p_3^2$ for even $n$ [Hall 1991]. |
| KierFlex | Kier molecular flexibility index: (KierA1) (KierA2) / $n$ [Hall 1991]. |
| logS | Log of the aqueous solubility This property is calculated from an atom contribution linear atom type model [Hou 2004] with $r^2$ = 0.90, ~1,200 molecules. |
| apol | Sum of the atomic polarizabilities (including implicit hydrogens) with polarizabilities taken from [CRC 1994]. |
| bpol | Sum of the absolute value of the difference between atomic polarizabilities of all bonded atoms in the molecule (including implicit hydrogens) with polarizabilities taken from [CRC 1994]. |
| | |
| PEOE_VSA_FHYD | Fractional hydrophobic van der Waals surface area. This is the sum of the $v_i$ such that $|q_i|$ is less than or equal to 0.2 divided by the total surface area. The $v_i$ are calculated using a connection table approximation. |
| PEOE_VSA_FNEG | Fractional negative van der Waals surface area. This is the sum of the $v_i$ such that $q_i$ is negative divided by the total surface area. The $v_i$ are calculated using a connection table approximation. |
| PEOE_VSA_FPNEG | Fractional negative polar van der Waals surface area. This is the sum of the $v_i$ such that $q_i$ is less than -0.2 divided by the total surface area. The $v_i$ are calculated using a connection table approximation. |
| PEOE_VSA_FPOL | Fractional polar van der Waals surface area. This is the sum of the $v_i$ such that $|q_i|$ is greater than 0.2 divided by the total surface area. The $v_i$ are calculated using a connection table approximation. |
| PEOE_VSA_FPOS | Fractional positive van der Waals surface area. This is the sum of the $v_i$ such that $q_i$ is non-negative divided by the total surface area. The $v_i$ are calculated using a connection table approximation. |
| PEOE_VSA_FPPOS | Fractional positive polar van der Waals surface area. This is the sum of the $v_i$ such that $q_i$ is greater than 0.2 divided by the total surface area. The $v_i$ are calculated using a connection table approximation. |
| PEOE_VSA_HYD | Total hydrophobic van der Waals surface area. This is the sum of the $v_i$ such that $|q_i|$ is less than or equal to 0.2. The $v_i$ are calculated using a connection table approximation. |
| PEOE_VSA_NEG | Total negative van der Waals surface area. This is the sum of the $v_i$ such that $q_i$ is negative. The $v_i$ are calculated using a connection table approximation. |
| PEOE_VSA_PNEG | Total negative polar van der Waals surface area. This is the sum of the $v_i$ such that $q_i$ is less than -0.2. The $v_i$ are calculated using a connection table approximation. |
| PEOE_VSA_POL | Total polar van der Waals surface area. This is the sum of the $v_i$ such that $|q_i|$ is greater than 0.2. The $v_i$ are calculated using a connection table approximation. |
| PEOE_VSA_POS | Total positive van der Waals surface area. This is the sum of the $v_i$ such that $q_i$ is non-negative. The $v_i$ are calculated using a connection table approximation. |
| PEOE_VSA_PPOS | Total positive polar van der Waals surface area. This is the sum of the $v_i$ such that $q_i$ is greater than 0.2. The $v_i$ are calculated using a connection table approximation. |
| vsa_other | Approximation to the sum of VDW surface areas of atoms typed as "other". |
| vsa_pol | Approximation to the sum of VDW surface areas of polar atoms (atoms that are both hydrogen bond donors and acceptors), such as -OH. |
| SlogP | Log of the octanol/water partition coefficient (including implicit hydrogens). This property is an atomic contribution model [Crippen 1999] that calculates logP from the given structure; i.e., the correct protonation state (washed structures). Results may vary from the logP(o/w) descriptor. The training set for SlogP was ~7000 structures. |
| SMR | Molecular refractivity (including implicit hydrogens). This property is an atomic contribution model [Crippen 1999] that assumes the correct protonation state |

<u>Support Vector Regression Algorithm</u>

The Support Vector Regression (SVR) algorithm is a supervised learning algorithm built based on the Support Vector Machine algorithm that's used for classification models. The SVR algorithm functions by finding the best optimal hyperplane (decision boundary) that maximizes the margin (ε), which is the distance between the closest support vectors on each side and the hyperplane, while minimizing the error between the predicted and actual values (Figure 7).

The SVR model predicts a continuous output variable by mapping the input data to a high-dimensional feature space using a kernel function and solving a convex optimization problem which involves adjusting the weights assigned to each feature in the input data, as well as the bias term to minimize the error between the predicted and actual output. Once the SVR model is trained, it can be used to predict the output values for new input data. The predicted values are calculated based on the weights and bias terms that define the hyperplane. The optimization problem is mathematically formulated by:

Minimize:

$$\frac{1}{2} \left|\left| \boldsymbol{w} \right|\right|^2 + C \sum_{i=1}^{n} |\xi_i|$$

Constraints:

$$|y_i - w_i x_i| \leq \varepsilon + |\xi_i|$$

Where **W** is the coefficient vector which corresponds to the weights assigned to the input vectors. *C* is the regularization hyperparameter that controls the trade-off between the margin ($\varepsilon$) and the absolute error $\left( |y_i - w_i x_i| \right)$. $\xi_i$ denotes a deviation from the margin.

The SVR algorithm tries to find a hyperplane (a nonlinear manifold in the kernel space) that has the widest margin possible to allow for some degree of error and prevent overfitting to the training data. Trying to find the best possible fit to the data while minimizing the coefficient vector (**W**) is essential to finding simpler models. Minimized coefficient vector corresponds to a reduced magnitude of weights assigned to input features which reduces the number of relevant features that has an influence on the model resulting in a simpler, easier to interpret model that's less prone to overfitting. While other kernels like the Radial basis function (RBF) and polynomial kernel functions by explicitly mapping the input data to a high-dimensional feature space first to make its linear calculation (Figure 8), in this experiment, the Python Scikit-learn SVR's Linear kernel was used. The Linear kernel maps the input data to a high-dimensional feature space with the same dimensionality as the original feature space using a dot product between data points. The Linear kernel creates this map without significant computational cost by using a sparse representation of the input vectors which includes only the non-zero input features. Solving an optimization problem with fewer support vectors (data points with non-zero coefficients) is computationally less demanding than solving one with a larger number of support vectors.

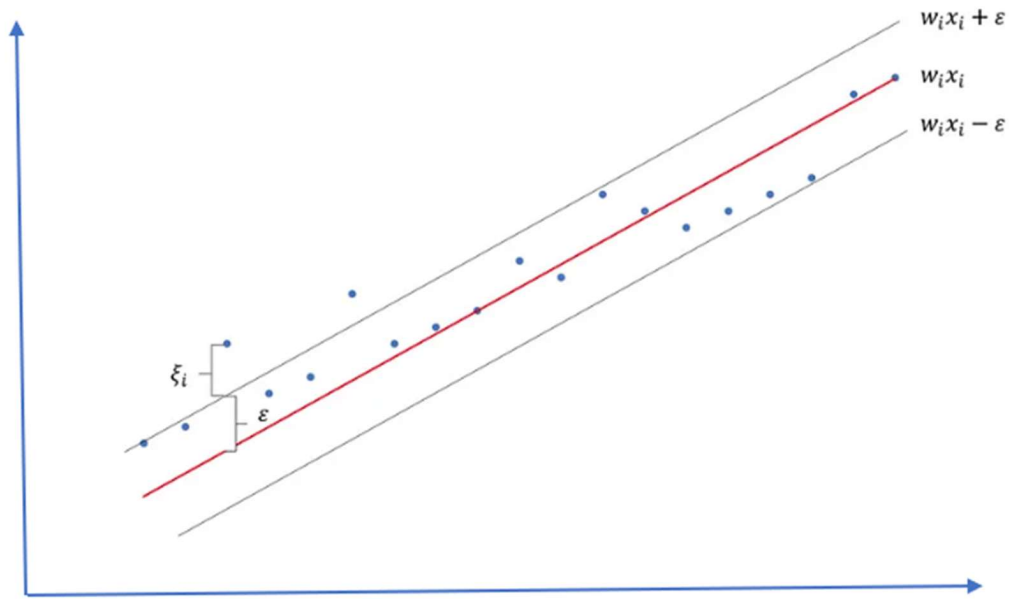

**Figure 8.** Graph illustration of the SVR mathematical algorithm.

Using the SVR algorithm to build models, the 2431 x 87 matrix dataset was split into 80% training set (1945 siRNAs and corresponding features) and 20% test set (486 siRNAs and corresponding features). A model was generated from fitting the training data into the SVR algorithm by calling the "svr.fit" function in python Scikit-learn package. Using a For loop, the data was randomly split into different 500 combinations of training and test sets for cross validation. 500 different SVR models were built from the different partitions and tested on their respective test sets. The predictive power of each model was measured by calculating the Root Mean Square Error (RMSE) of the model and the Pearson correlation coefficient-squared value ($r^2$) of the actual and predicted output for both training and test sets. Statistical analysis of the $r^2$ value of 500 the models was performed to show the distribution of the predictive models. The best model was used in the prediction of the potency of potential siRNAs against SARS-CoV-2 Spike gene.

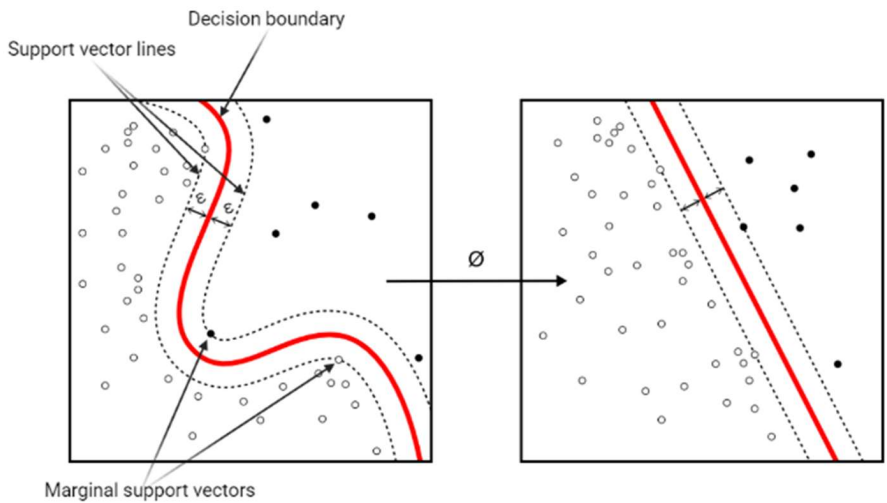

**Figure 9.** Schematic illustration of the process of mapping non-linear input vectors to a high-dimensional feature space to fit data with a linear regression model. The asymptotic notation (∅) describes the time complexity and computationally expensive nature of kernels (polynomial, RBF) that employ this method. Source: Wiki-SVM

Partial Least Squares Regression

Partial Least Squares (PLS) regression algorithm, like the ordinary least square (OLS) method, is a statistical method applied for multivariate or regression analysis. In the context of predictions, the PLS method is employed as a machine learning method where dependent variables are numerical. Both methods model the relationship between a dependent variable and multiple independent variables but unlike the OLS method, the PLS method supports multicollinearity of the independent variables by reducing the dimensionality of the correlated features while focusing on covariance.

The PLS regression method aims to construct latent variables from the independent variables by finding the linear combination of independent (predictor) variables that explains the maximum covariance with the dependent (response) variable. This helps to reduce the dimensionality of the data and improve the predictive power of the model. By focusing on the covariance between the predictor and response variables, PLS regression can identify the most important features in the data that are relevant for predicting the response variable. The PLS algorithm discovers the components by decomposing the predictor variables and the response variable into orthogonal subspaces, where the subspaces are defined by the scores of the previous components.

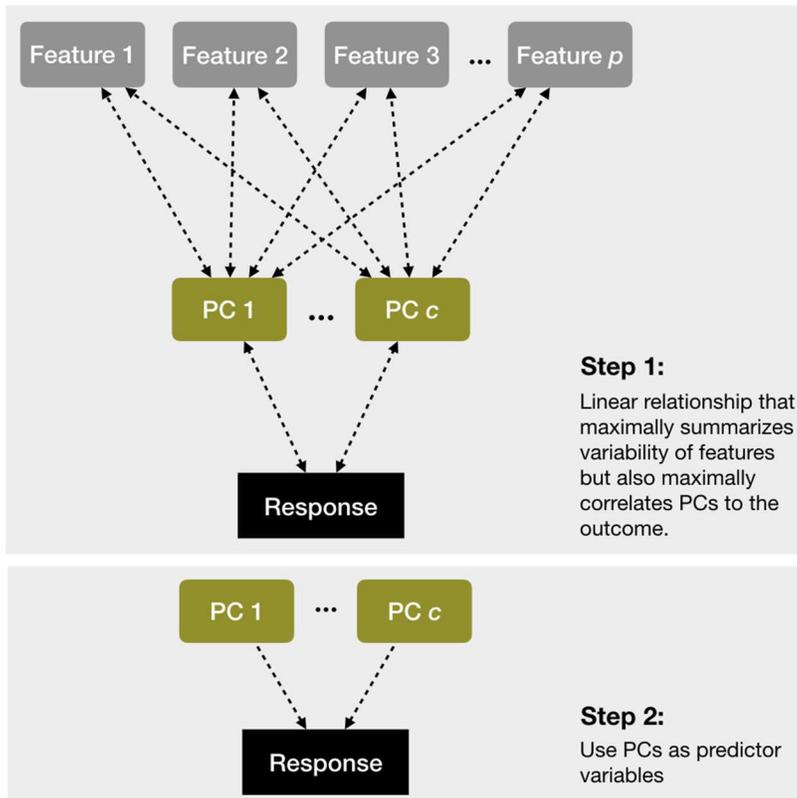

**Figure 10.** Schematic illustration of the PLS method. Source: <u>Supervised learning</u>

Mathematically. The following steps describes the process of the PLS regression:

1. Center and scale the predictor variables and the response variable to have zero mean and unit variance.

2. Initialize the first component by computing the linear combination of the predictor variables that has the highest covariance with the response variable.

3. Compute the scores of the predictor variables and the response variable for the first component.

4. Compute the loadings of the predictor variables and the response variable for the first component.

5. Orthogonalize the predictor variables and the response variable with respect to the scores of the first component.

6. Repeat steps 2-5 to find the second, third, and so on components, until the desired number of components is reached.

7. Use the scores of the components as the new predictor variables in a linear regression model to predict the response variable.

The optimization problem is formulated by:

Maximize:

$$\text{Cov}(Xw, Yc)$$

Constraint:

$$\|w\| = \|c\| = 1$$

where X is the matrix of independent variables, Y is the vector of dependent variable, w is the weight vector for the predictor variables, c is the weight scalar for the response variable, and Cov (Xw, Yc) is the covariance between the linear combinations of X and Y defined by w and c, respectively. This problem is solved iteratively for each component. In this experiment 3 components were used

Potency prediction of siRNAs against Sars-CoV-2 Spike gene.

To generate potential siRNAs against Sars-CoV-2 Spike gene, the Sars-CoV-2 Spike cDNA was sourced from NCBI. Using a DNA template conversion function in python, the Spike cDNA was converted to Spike mRNA. The Spike mRNA was 3822 nucleotide long. A programming function was written in python to generate all the potential siRNA substrings (21-nucleotide long) from the Spike gene mRNA. The 4 principal components generated from the PCA of 206 molecular descriptors of each nucleotide were incorporated to define the features of each potential siRNA sequence against Spike gene mRNA. The best SVR and PLS model was applied to predict the potency of all the potential siRNAs.

<u>Use Case of the Pre-trained Models – Application to SARS-CoV-2 Spike mRNA (Model Validation)</u>

To further validate the models, the best SVR and PLS models were used to predict the potency of previously experimentally validated siRNAs against Spike gene mRNA from Gallicano and colleagues' recent experiment on "Molecular targeting of vulnerable RNA sequences in SARS CoV-2: identifying clinical feasibility" [68]. Their experiment showed a siRNA dose-dependent inhibition of Sars-CoV-2 Spike protein in HEK293 and a primary human tracheal cell line. The models built were tested on the nucleotide sequence composition of a clinically feasible siRNA. See supplemental data for details on the codes (Appendix).

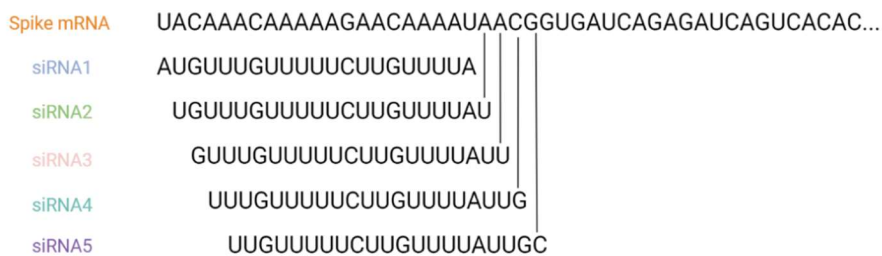

**Figure 11.** SARS-CoV-2 siRNA design. Schematic diagram of the selection of potential siRNA sequence against SARS-Cov-2 Spike mRNA.

CHAPTER IV

RESULTS

**Table 3**. Principal components of 206 molecular descriptors of each nucleotide

| Nucleotide | PC1 | PC2 | PC3 | PC4 |
|------------|-----|-----|-----|-----|
| "A" | -1.0349905 | -0.4186424 | -1.6574864 | 0.07919433 |
| "C" | 0.34019661 | 1.0113487 | 0.054668784 | -1.6906955 |
| "G" | -1.2724621 | -0.34483245 | 1.4934666 | 0.17744938 |
| "U" | 0.62386912 | 1.2274109 | -0.026959596 | 1.4503527 |
| "T" | 1.3433869 | -1.4752848 | 0.13631068 | -0.016300851 |

The principal component analysis (PCA) of 206 molecular descriptors (Table 2) of each nucleotide yielded 4 components.

<u>SVR Models</u>

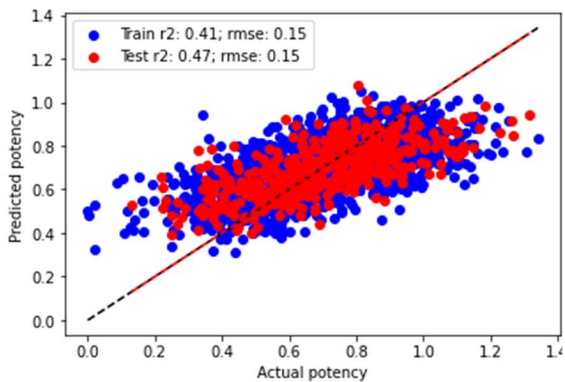
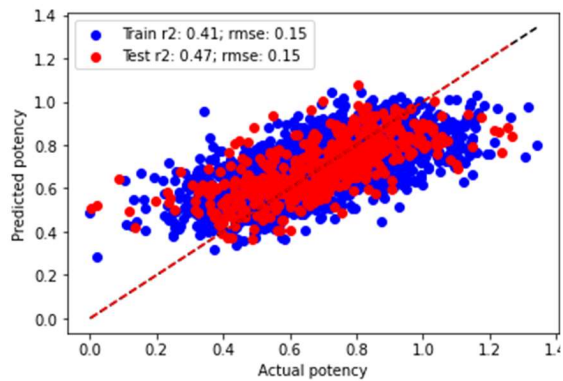

**Figure 12.** SVR Model 1- Random split 133

**Figure 13.** SVR Model 2- Random split 60

Figure 12 shows a correlation plot of the actual potency and the model predicted potency for both 80% training set and 20% test set split. The random split number describes the random state at which the data was partitioned. At random split number 133, the SVR algorithm presented the best fit for the data with a Pearson correlation coefficient-squared, $r^2$ value of 0.468 approximately 0.47 ($r = 0.68$) for the test set and 0.411($r = 0.64$) for the training set. The diagonal line represents the regression line with the red line representing the regression line of the test set and the black line representing the regression line for the training set. Each dot represents the potency values with the red dot representing the test set and the blue dot representing the training set.

Figure 13 shows a correlation plot of the actual potency and the model predicted potency for both 80% training set and 20% test set split with random split number 60. At the random split number 60, the SVR algorithm presented the next best fit for the data with a $r^2$ value of 0.467 approximately 0.47 ($r = 0.68$) for the test set and 0.4109 approximately 0.411 ($r = 0.64$) for the training set.

 Figure 14 shows a correlation plot of the actual potency and the model predicted potency for both 80% training set and 20% test set split with random split number 83. At random split number 83, the SVR algorithm presented a $r^2$ value of 0.462 ($r = 0.68$) and 0.413 ($r = 0.64$) for the test and training set respectively.

Figure 15 shows a correlation plot of the actual potency and the model predicted potency for both 80% training set and 20% test set split with random split number 20. At the random split number 20, the SVR algorithm presented a $r^2$ value of 0.461 ($r = 0.68$) and 0.415 ($r = 0.64$) for the test and training set respectively.

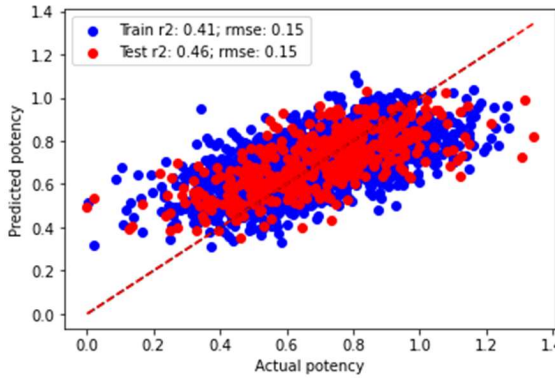 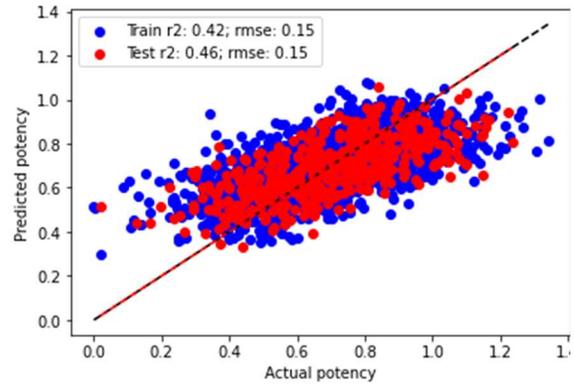

**Figure 14.** SVR Model 3- Random split 83

**Figure 15.** SVR Model 4- Random split 20

**Table 4.** Statistics of Pearson correlation coefficient-squared, $r^2$ of 500 SVR models

| Statistics of 500 SVR models Pearson correlation coefficient-squared, $r^2$ values | | | | |
|---|---|---|---|---|
| Dataset | Mean | Standard deviation | Maximum | Minimum |
| Training set | 0.4328 | 0.00699 | 0.4545 | 0.4109 |
| Test set | 0.3936 | 0.02922 | 0.4792 | 0.274 |

The average $r^2$ value for the correlation between the actual and predicted potency of 500 test sets was 0.3936 ($r = 0.627$) with a standard deviation of 0.02922. The average $r^2$ value for the correlation between the actual and predicted potency of 500 training sets was 0.4328 ($r = 0.658$) with a standard deviation of 0.00699.

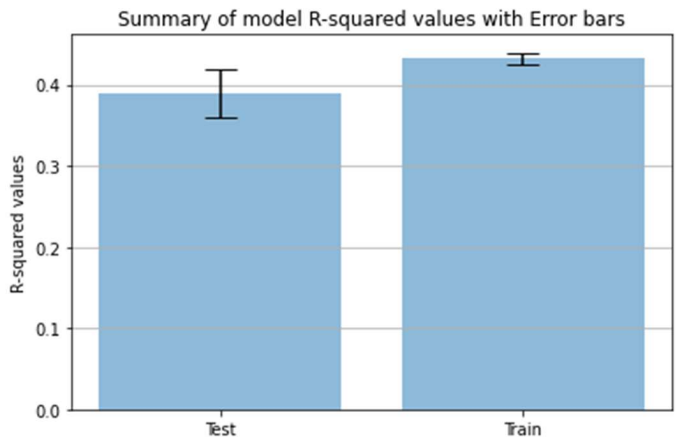

**Figure 16.** Error bars of 500 SVR models' Pearson correlation coefficient-squared, $r^2$ of both test and train dataset.

The error bars represent the standard deviation of the mean for each $r^2$ value as previously shown in Table 4.

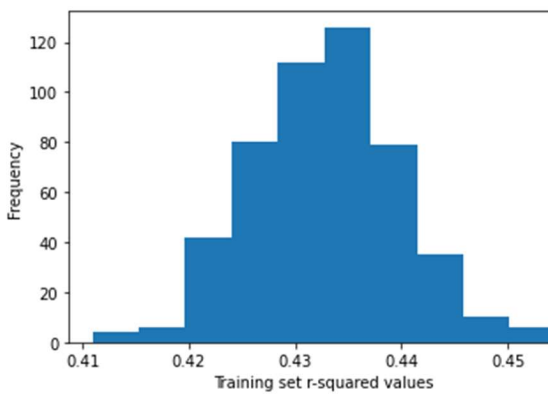

**Figure 17.** Histogram of 500 SVR models' $r^2$ values of the Training dataset.

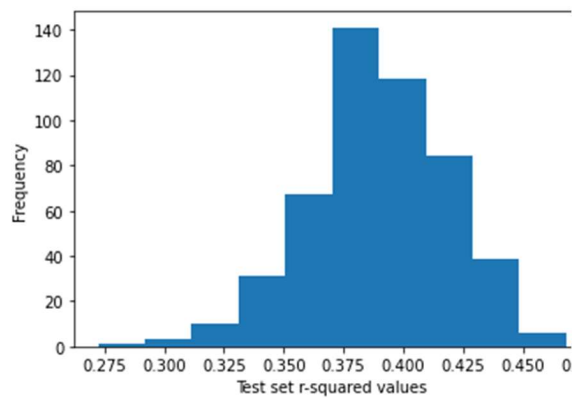

**Figure 18.** Histogram of 500 SVR models' $r^2$ values of the Test dataset.

<u>PLS Models</u>

Figure 19 shows a correlation plot of the actual potency and the model predicted potency for both 80% training set and 20% test set split. Like the SVR algorithm, the random split number describes the random state at which the data was partitioned. At random split number 60, the PLS algorithm presented the best fit for the data with a $r^2$ value of 0.479 approximately 0.48 ($r = 0.69$) for the test set and 0.415 approximately 0.42 ($r = 0.64$) for the training set.

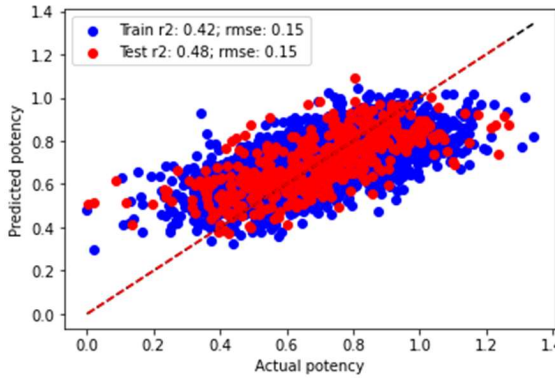 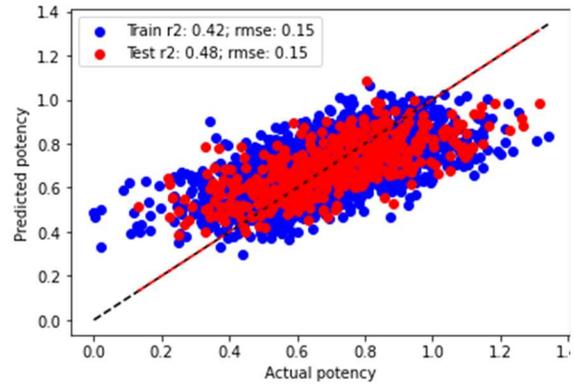

**Figure 19.** PLS Model 1- Random split 60

**Figure 20.** PLS Model 2-Random split 133

Figure 20 shows a correlation plot of the actual potency and the model predicted potency for both 80% training set and 20% test set split with random split number 133. At the random split number 133, the PLS algorithm presented the next best fit for the data with a $r^2$ value of 0.478 approximately 0.48 ($r = 0.69$) for the test set and 0.4165 approximately 0.42 ($r = 0.65$) for the training set.

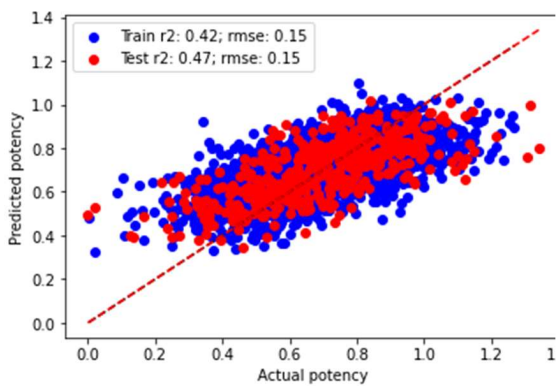 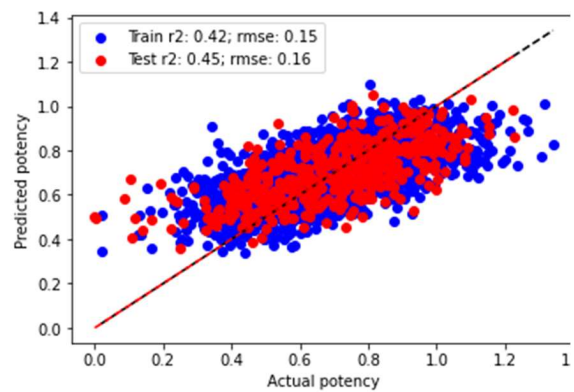

**Figure 21.** PLS Model 3-Random split 83    **Figure 22.** PLS Model 4-Random split 34

Figure 21 shows a correlation plot of the actual potency and the model predicted potency for both 80% training set and 20% test set split with random split number 83. At random split number 83, the PLS algorithm presented a $r^2$ value of 0.4719 ($r = 0.69$) and 0.4170 ($r = 0.65$) for the test and training set respectively.

Figure 22 shows a correlation plot of the actual potency and the model predicted potency for both 80% training set and 20% test set split with random split number 34. At random split number 34, the PLS algorithm presented a Pearson correlation coefficient-squared value of 0.4547 ($r = 0.67$) and 0.4208 ($r = 0.65$) for the test and training set respectively.

**Table 5.** Statistics of Pearson correlation coefficient-squared, $r^2$ of 500 PLS models

| Statistics of 500 PLS models Pearson correlation coefficient-squared, $r^2$ values | | | | |
|---|---|---|---|---|
| Dataset | Mean | Standard deviation | Maximum | Minimum |
| Training set | 0.4362 | 0.00681 | 0.4576 | 0.4153 |
| Test set | 0.3936 | 0.02942 | 0.4792 | 0.274 |

The average $r^2$ value for the correlation between the actual and predicted potency of 500 test sets was 0.3936 ($r = 0.627$) with a standard deviation of 0.02942. The average $r^2$ value for the correlation between the actual and predicted potency of 500 training sets was 0.4362 ($r = 0.66$) with a standard deviation of 0.00681.

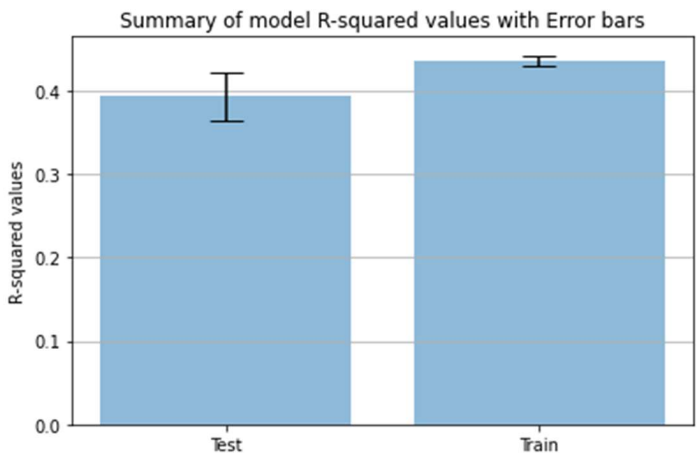

**Figure 23.** Error bars of 500 PLS models' $r^2$ values of both test and train dataset

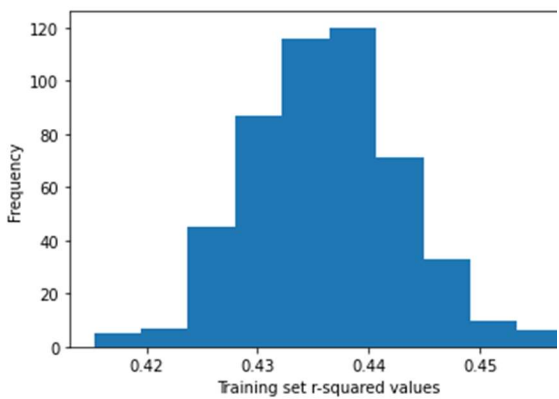

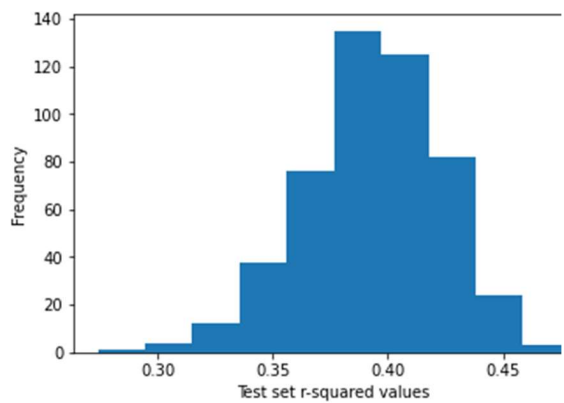

**Figure 24**: Histogram of 500 PLS models' *r²* values of the Training dataset

**Figure 25**: Histogram of 500 PLS models' *r²* values of the Test dataset

Potency prediction of siRNAs against Sars-CoV-2 Spike gene.

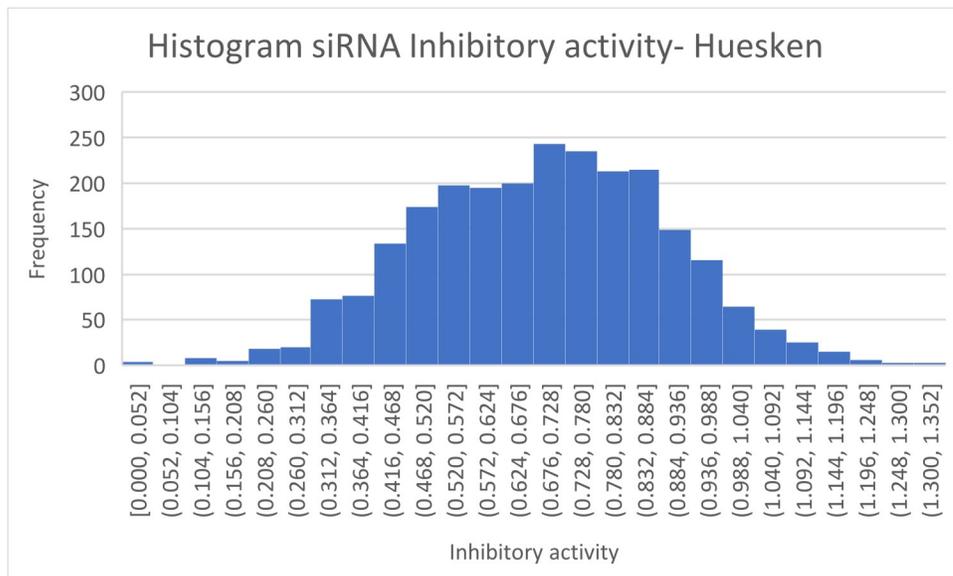

**Figure 26.** A histogram of the 2,431 siRNAs inhibitory activity, Huesken's dataset.

Authors of the Huesken dataset [91] quantitatively evaluated the inhibitory activity of siRNAs in the initial screening. The inhibitory activity values were normalized by the mean and standard deviation of the entire dataset before being used as input for the artificial neural network (ANN) during training. This allowed the ANN to distinguish between high and low inhibitory activity based on the standard deviations from the mean, rather than absolute % luciferase inhibition activity values. The histogram above shows the normalized inhibitory activity of corresponding siRNA at 50nM concentration.

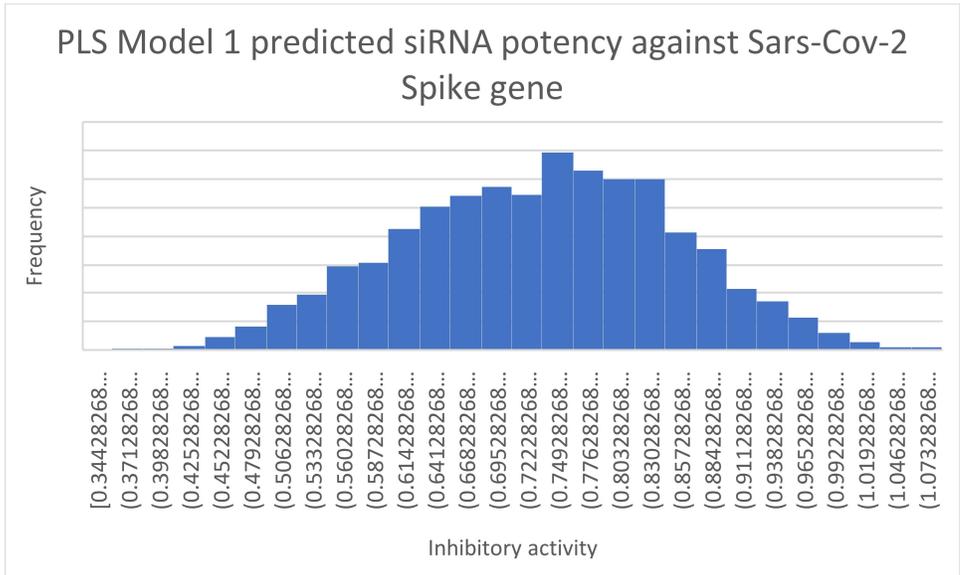

**Figure 27.** A histogram of the 3,802 siRNAs predicted inhibitory activity against SARS-CoV-2 spike mRNA using PLS model 1. Describes the normal distribution of the siRNA potency predicted by the best PLS model.

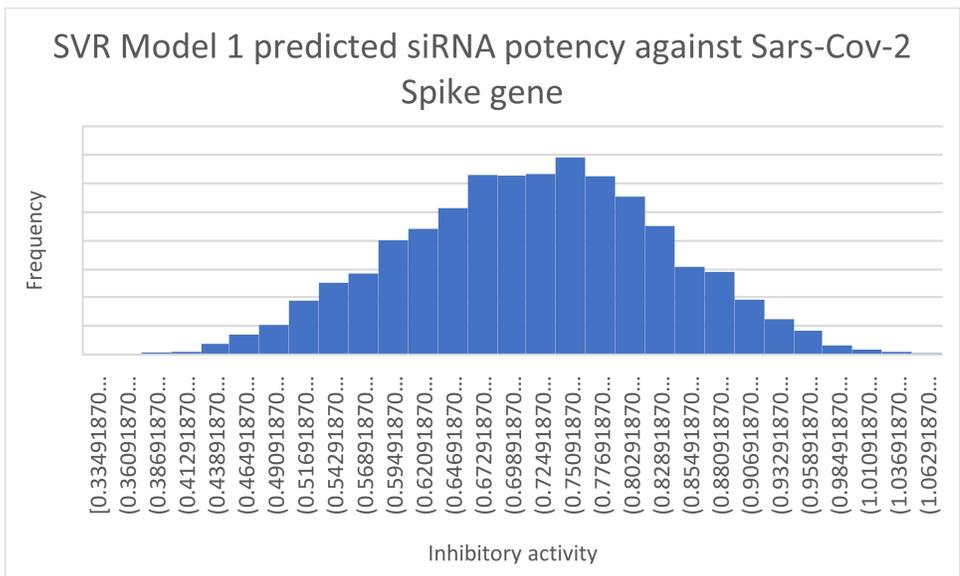

**Figure 28.** A histogram of the 3,802 siRNAs predicted inhibitory activity against SARS-CoV-2 spike mRNA using SVR model 1. Describes the normal distribution of the siRNA potency predicted by the best SVR model.

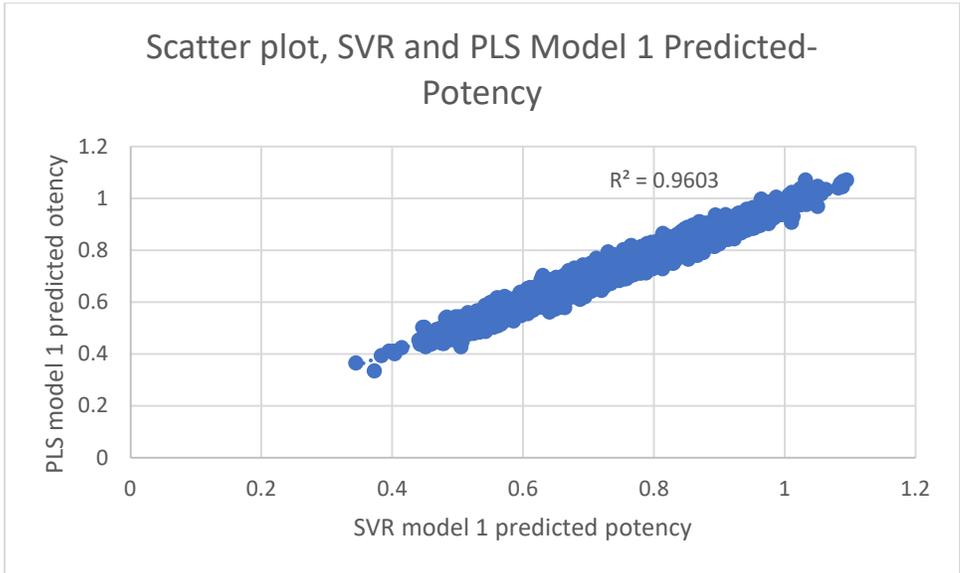

**Figure 29.** Comparison between SVR-predicted vs PLS-predicted potencies of 3,802 siRNAs against SARS-CoV-2 spike mRNA. The comparison of the best SVR and PLS models trained on the same data yields, $R^2 = 0.9603$ i.e., Pearson correlation coefficient, $r$ = 0.98.

**Table 6.** The siRNA represented in the table below had the best potency values after prediction with the best PLS model.

**Top 5 Highest Predicted potency using PLS model 1**

| ID | siRNA sequence | Predicted potency | sequence modification | Predicted potency |
|---|---|---|---|---|
| 829 | UAAAAUAUAAUGAAAAUGGAA | 1.094233 | UAAAAUAUAAUGAAAAUGGtt | 1.150986 |
| 397 | UUCAAUUUUGUAAUGAUCCAU | 1.089277 | UUCAAUUUUGUAAUGAUCCtt | 1.115319 |
| 3669 | UUGAUUGCCAUAGUAAUGGUG | 1.088174 | UUGAUUGCCAUAGUAAUGGtt | 1.084854 |
| 3466 | UUAAGAAUCAUACAUCACCAG | 1.085283 | UUAAGAAUCAUACAUCACCtt | 1.09949 |
| 2274 | UUUUGUACACAAUUAAACCGU | 1.081664 | UUUUGUACACAAUUAAACCtt | 1.10183 |

**Table 7.** The siRNA represented in the table above had the best potency values after prediction with the best SVR model.

**Top 5 Highest Predicted potency using SVR model 1**

| ID | siRNA sequence | Predicted potency | sequence modification | Predicted potency |
|---|---|---|---|---|
| 829 | UAAAAUAUAAUGAAAAUGGAA | 1.070515511 | UAAAAUAUAAUGAAAAUGGtt | 1.11186391 |
| 1543 | UUGAACUUCUACAUGCACCAG | 1.070242366 | UUGAACUUCUACAUGCACCtt | 1.08225311 |
| 397 | UUCAAUUUUGUAAUGAUCCAU | 1.0662553 | UUCAAUUUUGUAAUGAUCCtt | 1.11000159 |
| 3466 | UUAAGAAUCAUACAUCACCAG | 1.056909735 | UUAAGAAUCAUACAUCACCtt | 1.06892048 |
| 3409 | UUUAUGAUCCUUUGCAACCUG | 1.045656342 | UUUAUGAUCCUUUGCAACCtt | 1.05185638 |

Though no threshold was reported in the original Huesken data to determine higher or lower siRNA inhibitory activity, The siRNA sequences with the highest predicted inhibitory activity values were considered the most effective, while those with lower predicted inhibitory activity values were considered less effective.

<u>Use-Case of Pre-trained Models on Experimentally Validated SARS-CoV-2 siRNAs</u>

Ian Gallicano and colleagues [68] experimentally validated siRNA sequence against SARS-CoV-2 spike mRNA:

siRNA Antisense sequence - 'AUAAGUAGGGACUGGGUCUUU'

Using the best PLS and SVR models, the potency of the above siRNA was 0.84 and 0.81 respectively.

CHAPTER V

DISCUSSION

As a data pre-processing step, prior to machine learning modeling, feature scaling is recommended to normalize input data and prevent the model training being dominated by larger values while others are ignored which leads to biased models with poor performance. Feature scaling ensures all input features are given equal level of importance in the learning process. During this experiment, additional attempts to improve the model by scaling features didn't change the predictive power of the models because the input variables were principal components from PCA analysis.

The Pearson correlation coefficients-squared (coefficient of determination, $r^2$) values for both testing and training sets determined by comparing the actual potency values to the potency values predicted by the model and the RMSE values describes a method for evaluating each model. $r^2$ is a measure of how well the model fits the data, and it represents the proportion of the variance in the dependent variable that is explained by the independent variables. $r^2$ can range from 0 to 1, with higher values indicating better model performance. Evaluating the $r^2$ values of both test and training set within each model describes how the model is optimized to fit the training data compared to the generalization performance of the model on the test set.

The best SVR model with random split number 133 had a test $r^2$ value of 0.47 (Pearson correlation coefficient, $r = 0.69$) and a training $r^2$ value of 0.41(Pearson correlation coefficient, r = 0.64). The model with a comparable $r^2$ for both test and training set shows that the model's fit to the training data can generalize well on new unseen test set data. SVR model 2 with random split number 60 didn't differ from SVR model 1 with random split number 133, as such, the predictive power of the best SVR model is equal to SVR model 2.

Andrew Peek in a publication "Improving model predictions for RNA interference activities that use support vector machine regression by combining and filtering features" [88] built several models using the SVR algorithm to determine which features correlate with RNAi and the significance of feature filtering on model performance. Using the same 2431 siRNAs obtained from the Huesken dataset [91], Peek mapped each siRNA sequence to different numerical vector spaces. 14 different combinations of features ere mapped to the siRNAs based on specific nucleotide position composition, thermodynamics, entropy, guide strand structure and features, position independent N-gram, directional and non-directional target secondary structure, and target imprecise thermodynamics. Combining the numerical vectors of specific nucleotide position composition and N-grams features (1,444 feature number, FN) as inputs of the SVR model using a RBF kernel where model was trained and tested on the 2431 Huesken dataset and where the model was trained on the 2431 dataset and tested on 579 siRNA sequences from [98] shows a Pearson correlation coefficient, , $r = 0.783$ and $r = 0.492$ respectively. Several factors could account for the variation in Peek's results and the best SVR model with random split 133. The 579-siRNA test set was composed of 19 nucleotide long siRNAs. The number of input features was significant to the model's correlation coefficients. Larger numbers of input features gave better models [88]. At a more comparable number of features (Peek's FN = 91), $r = 0.736$. Peek's model performed poorly on novel test dataset but the best SVR model in this experiment predicted accurately when tested on a novel siRNA sequence (Gallicano's SARS-CoV-2 siRNA).

Gallicano's experimentally validated and clinically relevant siRNA designed against SARS-CoV-2 Spike gene [68] which had a dose-dependent inhibition of Spike mRNA was predicted to have a good potency values of 0.81 and 0.84 using the best SVR and PLS models respectively. Hence, the models built can be applied to predict the potency of siRNAs for further clinical development.

Evaluating the $r^2$ values of both test and training set of the best PLS model with random split number 60 had a test $r^2$ value of 0.48 (Pearson correlation coefficient, $r = 0.69$) and a training $r^2$ value of 0.42(Pearson correlation coefficient, $r = 0.65$). The model with comparable $r^2$ for both test and training set shows that the model's fit to the training data can generalize well on new unseen test set data. PLS model 2 with random split number 133 didn't differ from the best PLS model (PLS model 1) with random split number 60, as such, the predictive power of the best PLS model is equal to PLS model 2. A positive correlation of the predicted siRNA potency against the spike gene using the best PLS and best SVR model, $r = 0.98$ affirms the similarity of both machine learning algorithms.

Zheng and Dong's PLS model on the same 2431 Huesken data using BCUT descriptors of the siRNA as the input features reported a Pearson correlation coefficient, $r$ above 0.60. This result is comparable to the best PLS model with random split number 60 ($r = 0.69$).

Huesken's BioPredsi model built using artificial neural network (ANN) on the Huesken data reported a Pearson correlation coefficient, $r = 0.66$ between experimental inhibition observed and the predicted ANN inhibition score. This result is comparable to the best SVR and PLS model built in this experiment ($r = 0.69$).

Validating the nearly perfect correlation, $r = 0.98$ between the SVR and PLS potency predictions of siRNAs against SARS-CoV-2 (Fig. 29), there are overlapping siRNA sequences (siRNA ID 829, 3466 and 397) among the siRNAs with the highest predicted potencies predicted by both SVR and PLS models as shown in Table 6 and Table 7. Compared to Gallicano's SARS-CoV-2 siRNA Antisense sequence ('AUAAGUAGGGACUGGGUCUUU'), the GC content of the siRNAs with highest predicted potencies (Table 6 and Table 7) are relatively lower. Correspondingly, the predicted potencies of these highly potent siRNAs against SARS-CoV-2 are higher (> 1.0) than the predicted potency of Gallicano's siRNA (0.84 and 0.81). From this observation, the presence of GC base pairs plays a role in determining the inhibitory potency of siRNA. This corroborates the existing rule of efficient siRNA design that functional siRNA does not include long GC stretches. To improve the stability and protect the siRNA from exonuclease degradation, non-complementary TT overhangs are usually added to the 3' ends of the antisense strands. The base pair composition of the overhangs contributes to the efficiency of the siRNA as seen in Table 6 and Table 7 where the modification of siRNA against SARS-CoV-2 with TT overhangs changes the predicted potency of the siRNAs.

CHAPTER VI

CONCLUSION

Comparing the predictive power of the SVR and PLS models in this experiment with Huesken's ANN model, Dong, and Zheng's BCUT model, and other experimental and computational efforts to predict siRNA potency, the cheminformatics molecular description of siRNA nucleotide composition provides a valid method to design and predict the efficiency of siRNAs. Through the application of pre-trained model to predict the potency of siRNAs against SARS-CoV-2, COVID-19 therapeutic discovery process becomes more efficient. siRNAs can also be ranked according to their predicted potencies to select and experimentally validate the highly potent siRNAs.

Overall, training machine learning algorithms on large biological and chemical data enable efficient and less costly identification of potential drug candidates with higher accuracy. Adopting machine learning principle in the earlier stages of drug development to identify potential safety and efficacy issues early on can mitigate experimental and developmental costs associated with failure in later stages.